\documentclass[pra,superscriptaddress,showpacs,papersize=a4paper,twocolumn,floatfix]
{revtex4-1}%
\usepackage{float}
\usepackage{graphicx}
\usepackage[usenames,dvipsnames]{color}
\usepackage{amsmath}
\usepackage{bm}
\usepackage{amsfonts}
\usepackage{amssymb}
\def\be{\begin{equation}}
\def\ee{\end{equation}}

\def\Tr{\mathop{\rm Tr}}

\renewcommand{\Re}{\mathop{\rm Re}}
\renewcommand{\Im}{\mathop{\rm Im}}
\newcommand{\eps}{\epsilon}

\begin{document}

\title{Microwave response of a superconductor beyond the Eliashberg theory}

\begin{abstract}
We review recent progress in the theory of electromagnetic response of dirty superconductors subject to microwave radiation. The theory originally developed by Eliashberg in 1970 and soon after that elaborated in a number of publications addressed the effect of superconductivity enhancement in the vicinity of the transition temperature. This effect originates from nonequilibrium redistribution of quasiparticles and requires a minimal microwave frequency depending on the inelastic relaxation rate and temperature. In a recent series of papers we generalized the Eliashberg theory to arbitrary temperatures $T$, microwave frequencies $\omega$, \emph{dc} supercurrent, and inelastic relaxation rates, assuming that the microwave power is weak enough and can be treated perturbatively. In the phase diagram ($\omega,T$) the region of superconductivity enhancement occupies a finite area located near $T_c$. At sufficiently high frequencies and low temperatures, the effect of direct depairing prevails over quasiparticle redistribution, always leading to superconductivity suppression.
\end{abstract}

\author{K. S. Tikhonov}
\affiliation{L. D. Landau Institute for Theoretical Physics,
Chernogolovka 142432, Russia}
\affiliation{Skolkovo Institute of Science and Technology, Moscow, 121205, Russia}
\affiliation{Institut f{\"u}r Nanotechnologie, Karlsruhe Institute of Technology, 76021 Karlsruhe, Germany}

\author{A. V. Semenov}
\affiliation{Physics Department, Moscow State University of Education, Moscow 119992, Russia}
\affiliation{Moscow Institute of Physics and Technology, Dolgoprudny, Moscow 141700,
Russia}

\author{I. A. Devyatov}
\email[]{Deceased on June 11th, 2019.}
\affiliation{Lomonosov Moscow State University, Skobeltsyn Institute of Nuclear
Physics, 1(2), Leninskie gory, GSP-1, Moscow 119991, Russia}
\affiliation{Moscow Institute of Physics and Technology, Dolgoprudny, Moscow 141700,
Russia}

\author{M. A. Skvortsov}
\affiliation{Skolkovo Institute of Science and Technology, Moscow, 121205, Russia}
\affiliation{L. D. Landau Institute for Theoretical Physics,
Chernogolovka 142432, Russia}

\maketitle

\section{Introduction}

Theoretical study of the depairing effect of a \emph{dc} current and \emph{dc} magnetic field started soon after the creation of the microscopic theory of superconductivity by Bardeen, Cooper and Schrieffer (BCS) \cite{bardeen1957theory}. It was shown \cite{anderson1958coherent} that a \emph{dc} current modifies the ground state of a superconductor, with the Cooper pairs acquiring a non-zero momentum. That results in the modification of the spectral properties: the value of superconducting order parameter is decreased, and the BCS singularity near the gap is smeared. The equivalence of depairing action of a \emph{dc} current and \emph{dc} magnetic field to that of paramagnetic impurities \cite{AGmagnetic} was demonstrated theoretically \cite{Maki-Parks} and proven experimentally \cite{anthore2003density}. For dirty superconductors, i.e. with the elastic mean free path much shorter than the BCS coherence length, the theory of depairing by a \emph{dc} current and \emph{dc} field was elaborated in Ref. \cite{kupryanov1980temperature} and experimentally verified in Refs.\ \cite{romijn1982critical, anthore2003density}.

In the vicinity of the critical temperature, the effect of a microwave field is mainly related to redistribution of quasiparticles. Remarkably, irradiation may lead not only to suppression but to enhancement \cite{dayem1967behavior,wyatt1966microwave} of superconductivity under certain conditions. The alteration of the superconducting gap due to a non-equilibrium distribution of quasiparticles created by a microwave field was theoretically explained by Eliashberg \cite{eliashberg70,ivlev1971influence} in the framework of Gor'kov equations \cite{gorkov1969superconducting}. 
Early results in the field were summarized in the review \cite{mooij1981enhancement}.

At the same time, the effect of a microwave field on the spectral properties of a superconductor, i.~e.\ the modification of its ground state, was up to a recent time in a shadow. Under experimental conditions available in 70-s, either the effects related to quasiparticles were dominant, or the modification of the spectral functions by the embedded microwave was too small to be observable. Theoretical description of the modification of the ground state by a microwave field was developed just recently \cite{semenov2016coherent,Tikhonov2018,moor2017amplitude,Semenov-2B}. It has been stimulated by a growing applied interest to the interaction between superconductors and microwave field at very low temperatures, when the number of thermal quasiparticles is vanishingly small and the microwave response is governed by the modification of spectral properties. This is the conditions of operation of many prospective low-temperature devices, including superconducting micro-resonators \cite{de2014evidence, de2014fluctuations}, parametric amplifiers \cite{eom2012wideband}, and kinetic-inductance microwave detectors \cite{zmuidzinas2012superconducting}. The fundamental side of the problem is related to the search for the Higgs mode in superconductors \cite{beck2013transient,matsunaga2013higgs,matsunaga2014light,Silaev2019}.

\begin{figure}
\centering
\includegraphics[width=0.44\textwidth]{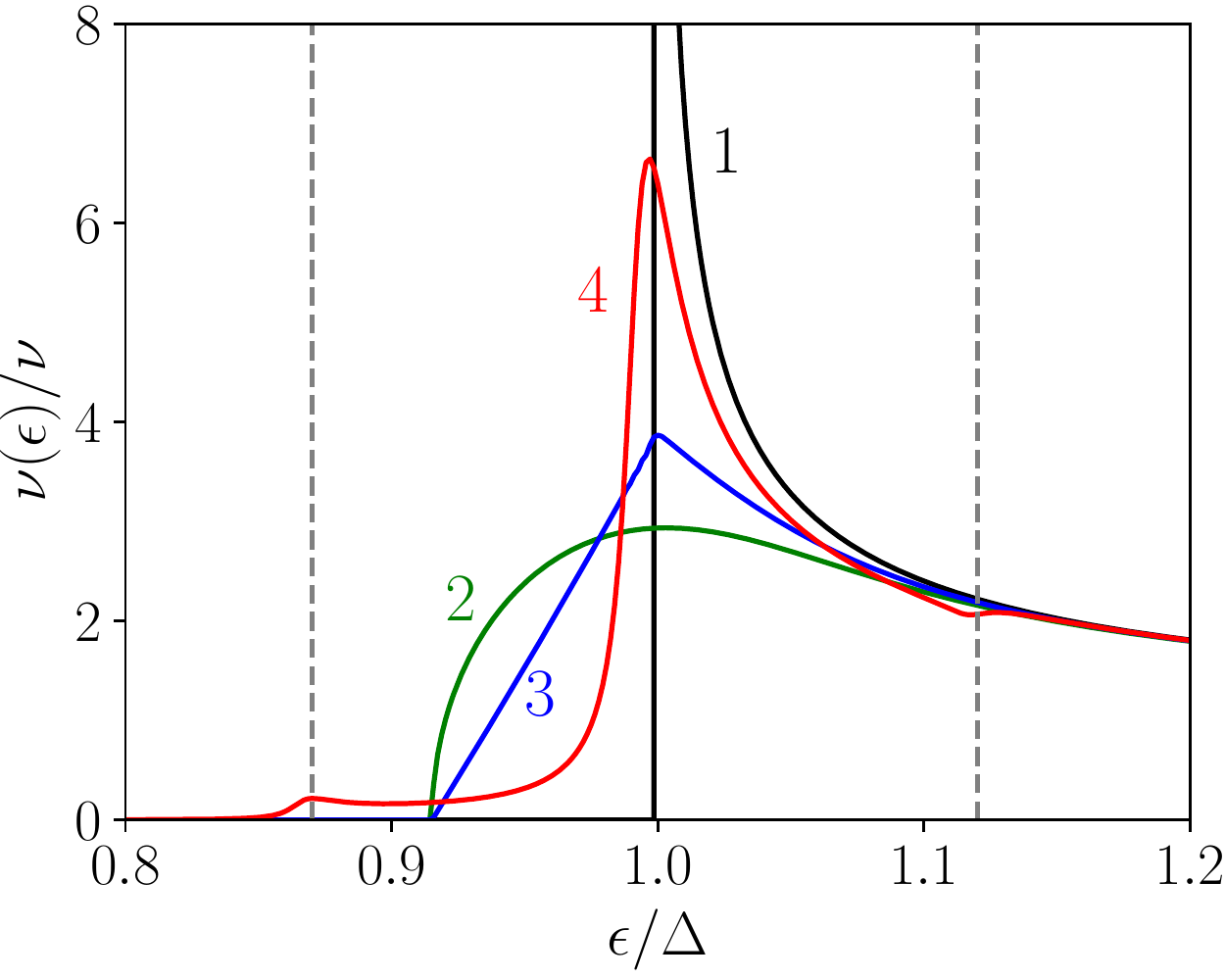}
\caption{Time-averaged density of states in a dirty superconductor at low temperature ($T\ll T_c$) under different biasing: (1) zero current (BCS), (2) \emph{dc} current, (3) low frequency current ($\hbar\omega\ll\alpha$), and (4) microwave current ($\alpha\ll\hbar\omega$). In all non-BCS cases, the value of the current or its amplitude is $0.25\,I_c$.}
\label{DOS}
\end{figure}

In this paper, we present a review of our recent results on extending the Eliashberg theory to the case of arbitrary temperatures and frequencies of the microwave field. We allow for an arbitrary \emph{dc} supercurrent and model the inelastic relaxation by escape to a reservoir.
Accounting for the effects of microwaves on both the spectral properties (direct depairing) and on the distribution of quasiparticles, we calculate the full phase diagram of a dirty superconductor and determine the regions of suppression/enhancement of the order parameter and of the critical current.

As we demonstrated in Ref.\ \cite{semenov2016coherent}, the spectral properties of a superconductor in a microwave field differ qualitatively from these of a superconductor with no current, with a \emph{dc} current or with a low-frequency current. This is illustrated by Fig.~\ref{DOS} for the density of states (DOS). The BCS peak is smeared, and additional features at `photon points' $\Delta \pm n\hbar\omega$ emerge. The latter can be understood in terms of the Floquet or quasienergy states \cite{zel1967quasienergy,grifoni1998driven}. In a microwave field, the eigenstates of electrons are not stationary states with definite energies, but the Floquet states with definite quasienergies. Being expanded in the energy basis, each Floquet state is a sum of components with energies differing by $\hbar\omega$. If $\hbar\omega$ is large compared to the relevant classical energy scale of the field $\alpha$ [see Eq.\ \eqref{Gamma} for the definition], this gives replicas of the BCS peak shifted to the `photon points'.
It was also shown that there appears an exponential-like tail of the DOS in the sub-gap region. All these predictions are in strong contrast with (i) the well-studied case of a \emph{dc} current \cite{kupryanov1980temperature, romijn1982critical, anthore2003density}, where the BCS peak is smeared without emergence of any additional peculiarities and a true gap in the DOS survives, and with (ii) the case of a low-frequency supercurrent ($\hbar\omega\ll\alpha$), where the DOS can be calculated as a time-average of DOS corresponding to the instant value of the current \cite{gurevich2014reduction}.

The Eliashberg theory explains the microwave-induced enhancement of superconductivity by redistribution of quasiparticles away from the superconducting gap due to absorption of $\hbar\omega$ quanta. That effectively cools electrons near the gap, which are responsible for pairing, leading to the increase of the order parameter and the spectral gap. An important ingredient of this mechanism is energy relaxation, which competes with the energy lift-up of quasiparticles and makes the nonequilibrium stationary. This competition sets a natural lower bound on microwave frequency, $\omega_\text{min}(T)$, at which the enhancement does exist \cite{eliashberg70}. Dependence of $\omega_\text{min}(T)$ on the inelastic relaxation rate can be used for experimental determination of the latter \cite{van1984inelastic}. Similar ideas have been discussed theoretically for superconducting weak links \cite{artemenko1979theory,schmid1980dynamic} and SNS junctions \cite{lempitskii1983stimulation,virtanen2011linear,tikhonov2015admittance}, and studied in recent experiments \cite{chiodi2011probing,dassonneville2013dissipation}.

\begin{figure}
\centering%
\includegraphics[width=0.47\textwidth]{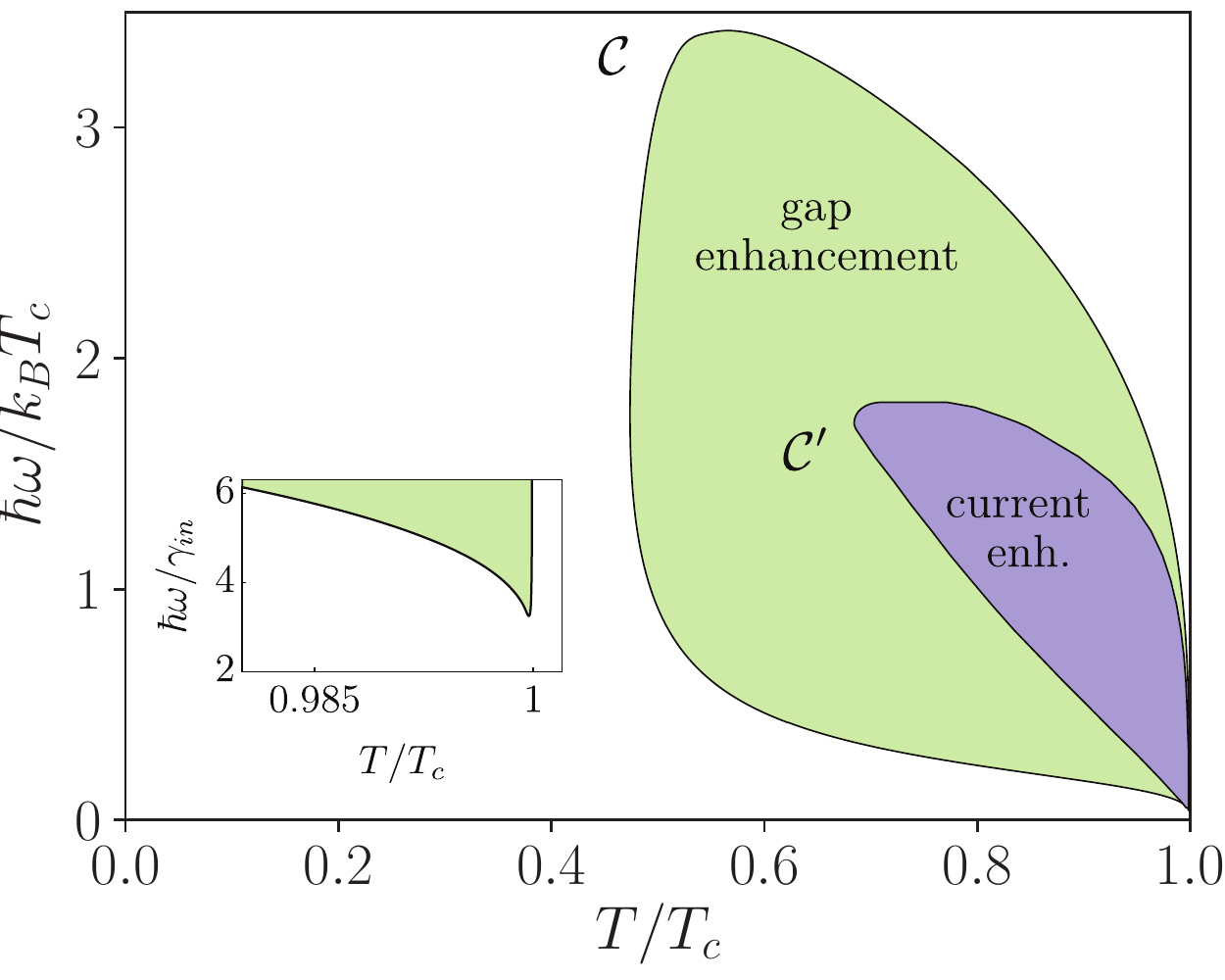}
\caption{Phase diagram of a dirty superconductor subject to a microwave radiation with vanishing power ($\alpha\to 0$). The gap is enhanced inside the region marked by the curve $\mathcal{C}$. The critical current is enhanced inside the region marked by the curve $\mathcal{C}'$. The inelastic relaxation is modeled by tunneling to a normal reservoir with the rate $\gamma_\text{in}/k_BT_c=0.02$. 
Inset: zoom of the gap enhancement region near $T_c$, showing the minimal frequency $\omega_\text{min,min}\approx
3.23 \, \gamma_\text{in}/\hbar$.}
\label{wmin-gap}
\end{figure}

Our results for enhancement and suppression of superconductivity by a weak microwave field ($\alpha\to0$) are summarized by the phase diagram in the $(\omega,T)$ plane shown in Fig.~\ref{wmin-gap}. 
In the absence of a \emph{dc} supercurrent, the region where the superconducting order parameter $\Delta(T)$ is enhanced compared to its zero-field value $\Delta_0(T)$ is located inside the contour $\mathcal{C}$. 
The upper bound $\omega_\text{max}(T)$ is set up by heating and the lower bound $\omega_\text{min}(T)$ is determined by inelastic relaxation.
 The bound from the side of low temperatures emerges due to the direct depairing by the induced microwave supercurrent.

One can also ask about the enhancement of superconductivity in terms of the critical current. The region where the critical current density is greater than the corresponding equilibrium value at a given temperature, is encircled by the contour $\mathcal{C'}$ in Fig.~\ref{wmin-gap}. It is narrower than the region of the order parameter enhancement. Indeed, it is harder to enhance superconductivity in the presence of a \emph{dc} supercurrent: the latter smears BCS singularity in the DOS \cite{Maki-Parks,anthore2003density} and the corresponding singularity in the nonequilibrium distribution function produced by the absorbed microwave field.

\section{Phase diagram of a superconductor in a microwave field}

\subsection{Model}

We consider a quasi-one-dimensional superconducting wire when both the \emph{dc} (supercurrent) and \emph{ac} (microwave field) components of the vector potential are parallel to the wire. Assuming the modulus of the order parameter be uniform along the wire \cite{eliashberg70,ivlev1971influence,schmid1977stability,eckern1979stability,schmid1980dynamic,van1984microwave,van1984inelastic}, we can gauge out the spatial dependence of the phase and work with a real order parameter subject to a time-dependent vector potential
\be
\label{A}
\mathbf{A}(t) = \mathbf{A}_{0}+\mathbf{A}_{1}\cos\omega t,
\ee
where the static part $\mathbf{A}_0$ accounts for the \emph{dc} supercurrent, and $\mathbf{A}_1{\parallel}\mathbf{A}_0$.
Both components of the vector potential act as pair breakers, and their effect can be characterizes by the energy scales \cite{Maki-Parks}
\be
\label{Gamma}
\Gamma = \frac{2e^{2}D\mathbf{A}_0^2}{\hbar c^2},
\qquad
\alpha = \frac{2e^{2}D\mathbf{A}_1^2}{\hbar c^2},
\ee
where $D$ is the normal-state diffusion coefficient in the superconductor \cite{defalpha}.
The depairing rate $\Gamma$ of a static supercurrent plays the role of spin-flip rate in the theory of magnetic impurities \cite{Maki-Parks}. It smears the BCS coherence peak, shifting the gap to $E_g=\Delta[1-(\Gamma/\Delta)^{2/3}]^{3/2}$ \cite{AGmagnetic}.

Below we discuss how to treat the problem of electromagnetic response in the lowest order in the microwave power $\alpha$ and arbitrary $T$, $\omega$, $\Gamma$, and inelastic width $\gamma_\text{in}$ \cite{gamma-com}, which will be modeled by quasiparticle tunneling to a reservoir.

\subsection{General scheme}

The response of a superconductor to microwave irradiation belongs to the class of most complicated problems in the theory of nonequilibrium superconductivity. Provided the material is far from the insulator transition, its theoretical description is based on dynamic equations for the quasiclassical Green's functions in the Keldysh representation. In the dirty limit, those are the Usadel equation for the Green's functions $\check g$, with the Keldysh component containing the kinetic equation for the distribution function \cite{larkin1986nonequilibrium,Kopnin-book}.
While the Green's function at equilibrium is diagonal in the energy space, $\check g_{\eps,\eps'} = 2\pi\delta(\eps-\eps)\check g_\eps$, the main difficulty associated with the nonequilibrium situation is the dependence of $\check g_{\eps,\eps'}$ on both energy arguments, that is a mathematical manifestation of transitions between the states at energies $\eps$ and $\eps\pm\omega$ induced by the microwave field. 

Since the nonlinear Usadel equation should be additionally supplemented by the self-consistency equation for the time-dependent order parameter, the resulting theory becomes too complicated to be treated analytically. It can be attacked either numerically (see e.g. Ref.\ \cite{SnymanNazarov}) or by perturbative analysis, assuming that the \emph{ac} component of the vector potential $\mathbf{A}_1(t)$ is small and can be treated as a perturbation on top of the steady state in the presence of a static $\mathbf{A}_0$. This is essentially the approximation utilized by Eliashberg \cite{eliashberg70} and in subsequent studies \cite{mooij1981enhancement} based on the Ginzburg-Landau (GL) expansion. Going beyond the GL region one has to work with the full set of the Usadel equations and to linearize the solution in the amplitude of the microwave radiation \cite{semenov2016coherent,moor2017amplitude,Semenov-2B}. However even in that case calculations are quite lengthy due to a nonlinear and nonlocal-in-time constraint imposed on $\check g$. 
In Ref.\ \cite{Tikhonov2018} we used a technically more convenient approach of the nonlinear Keldysh $\sigma$ model for superconducting systems \cite{FLS} to make a perturbative expansion in $\mathbf{A}_1(t)$. Both approaches are fully equivalent since the Usadel equation is nothing but the saddle point of the $\sigma$ model, but working with the latter gave us access to the standard machinery for expanding in terms of soft modes (diffusons and cooperons).

The Usadel equation is written for Green's function $\check g_{tt'}$ which bares two time (or energy) arguments and acts in the tensor product of the Nambu and Keldysh spaces, with the Pauli matrices $\tau_i$ and $\sigma_i$, respectively. In what follows we will consider time (or energy) arguments as usual matrix indices, with matrix multiplication implying convolution in the time (or energy) domain. The $\check g$ matrix satisfies the nonlinear constraint $\check g^2=1$.
In the zero-dimensional case (spatially uniform configurations), it obeys the Usadel equation
\begin{equation}
\label{action}
[i\epsilon\tau_3 - \Delta\tau_1 
- \hbar D \mathbf{a}\tau_{3} \check g \mathbf{a} \tau_{3}
+ \Sigma_\text{in}
, \check g ] = 0 ,
\end{equation}
where $\mathbf{a}(t) = e\mathbf{A}(t)/\hbar c$, and 
the order parameter $\Delta(t)$ should be determined from the self-consistency equation
\be
\label{SCE}
\Delta = - \frac{i\pi\lambda}{4} 
\Tr\bigl(\tau_1 \check g^K\bigr)
,
\ee
where $\lambda$ is the dimensionless Cooper coupling. 

The matrix $\Sigma_\text{in}$ describes inelastic relaxation, which can be due to electron-phonon interaction \cite{chang1977gap}, electron-electron interaction \cite{Pothier1997}, and escape to reservoirs. 
Qualitatively all three mechanisms have the same influence on the properties of the system.
To simplify the analysis we will assume that inelastic relaxation is dominated by tunneling to a reservoir, in which case
\be
\label{Sigma}
\Sigma_\text{in} 
= 
- \frac{\gamma_\textrm{in}}{2} \check g_\textrm{res} ,
\ee
where the escape rate $\gamma_\textrm{in}$ is proportional to the tunnel conductance of the interface. The matrix $\check g_\text{res}$ refers to the Green's function in the reservoir, which can be either normal or superconducting. Both cases were considered by the authors \cite{Tikhonov2018,DeviatovSemenov2019}, leading to very similar results for the microwave response. Therefore we focus here only on the case of a normal reservoir \cite{Tikhonov2018}. At equilibrium with the temperature $T$, its Green's function has the form
\be
\check g_\textrm{res}
=
\begin{pmatrix} 1 & 2F_0 \\ 0 & -1 \end{pmatrix}_\textrm{K} \otimes \tau_3 ,
\ee
where $F_0$ is diagonal in the energy representation, with $F_0(\epsilon)=1-2f_0(\epsilon)=\tanh(\epsilon/2T)$ being the thermal distribution function.

At equilibrium the Green's function $\check g$ is diagonal in the energy space, $\check g_{\epsilon\epsilon'} = 2\pi\delta(\epsilon-\epsilon')\check g(\epsilon)$, where \be
\label{Qsaddle}
\check g(\epsilon)
=
\begin{pmatrix} \hat g^R(\epsilon) & [\hat g^R(\epsilon)-\hat g^A(\epsilon)]F_0(\epsilon) \\ 0 & \hat g^A(\epsilon) \end{pmatrix}_\textrm{K} ,
\ee
with
\begin{subequations}
\begin{gather}
\hat g^R(\epsilon)
=
\begin{pmatrix}
\cos\theta^R(\epsilon) & \sin\theta^R(\epsilon) \\
\sin\theta^R(\epsilon) & - \cos\theta^R(\epsilon)
\end{pmatrix}_\textrm{N} ,
\\
\hat g^A(\epsilon)
=
-
\begin{pmatrix}
\cos\theta^A(\epsilon) & \sin\theta^A(\epsilon) \\
\sin\theta^A(\epsilon) & - \cos\theta^A(\epsilon)
\end{pmatrix}_\textrm{N} .
\end{gather}
\end{subequations}
The spectral angles obey the symmetry relations $\theta^{A}(\epsilon)=-\theta^{R}(-\epsilon)=-[\theta^{R}(\epsilon)]^{\ast}$ and can be found from Eq.\ \eqref{Sigma}, whose only energy-diagonal component reads
\be
\label{Usadel}
\Delta\cos\theta^{R}(\epsilon)+i\epsilon^{R}\sin\theta^{R}(\epsilon
)-\Gamma\sin\theta^{R}(\epsilon)\cos\theta^{R}(\epsilon)=0 .
\ee
Here $\epsilon^{R,A}=\epsilon\pm i\gamma_\textrm{in}/2$ and the depairing energy $\Gamma$ is defined in Eq.\ \eqref{Gamma}. The spectral angle obtained from Eq.\ \eqref{Usadel} for a given $\Delta$ should be substituted into the self-consistency equation \eqref{SCE}, which can be cast in the form
\begin{equation}
{\cal F}_\textrm{eq}(\Delta,\Gamma,T,\gamma_\textrm{in}) = 0,
\label{SCE-gen}
\end{equation}
where ${\cal F}_\textrm{eq}$ is defined as
\begin{equation}
\mathcal{F}_\textrm{eq}
=
\frac{1}{2\Delta} \int d\epsilon \,
F_0(\epsilon) \Im \sin\theta^R_\epsilon
- \frac{1}{\lambda}.
\label{F-eq}
\end{equation}

In the presence of a monochromatic radiation described by the vector potential (\ref{A}), the Green's function $\check g_{\eps,\eps'}$ acquires off-diagonal components in the energy space. For weak radiation power $\alpha$, linear-in-$\alpha$ corrections to the equilibrium Green's function (\ref{Qsaddle}) can be obtained perturbatively. In Ref.\ \cite{Tikhonov2018} this procedure was done at the level of the $\sigma$ model, where all possible sources of such a dependence were taken into account. The corresponding modification of the time-averaged spectral angles in the limit of a vanishing \emph{dc} current ($\Gamma=0$) and small temperatures ($T\ll T_c$) was discussed in Refs.\ \cite{semenov2016coherent,Semenov-2B}, where solution of the Usadel equation was treated in the first order in $\alpha$.

In order to develop a perturbative approach in the microwave power valid at arbitrary temperatures, one has to take into account that the critical temperature is shifted due to irradiation. As a result, in the vicinity of $T_c$ modification of the spectral angles becomes large and cannot be treated perturbatively. To overcome that obstacle, in Ref.\ \cite{Tikhonov2018} we suggested to use a scheme when the perturbative in $\alpha$ correction to the spectral function is calculated at a given order parameter $\Delta$ (different from the equilibrium value $\Delta_0$). The obtained correction is substituted then into the self-consistency equation (\ref{SCE}), where the first-order in $\alpha$ terms should be retained. Such an approach is in line with the GL derivation when quasiparticle degrees of freedom are integrated out to get the effective free energy of the order parameter field, but extends it to the case of arbitrary temperatures.

The resulting equation for the time-averaged order parameter that generalizes Eq.\ \eqref{F-eq} to the nonequilibrium case can then be written in the form
\begin{equation}
{\cal F}_\textrm{eq}(\Delta,\Gamma,T,\gamma_\textrm{in}) + \alpha {\cal F}_\textrm{neq}(\Delta,\Gamma,T,\omega,\gamma_\textrm{in}) = 0 ,
\label{SCE-gen}
\end{equation}
where the last term is just the first perturbative correction in $\alpha$ obtained as discussed above. Equation (\ref{SCE-gen}) should be used in order to determine the regions of enhancement/suppression of the order parameter by microwaves. The analytic expression for $F_\textrm{neq}$ is very lengthy and will be presented below only in some limiting cases. In general, ${\cal F}_\textrm{neq}(\Delta,\Gamma,T,\omega,\gamma_\textrm{in})$ should be calculated numerically.

\subsection{Zero-current case}
\label{SS:I0}

A peculiarity of the situation in the absence of a \emph{dc} supercurrent ($\Gamma=0$), is that the non-equilibrium correction ${\cal F}_\textrm{neq}(\Delta,\Gamma,T,\omega,\gamma_\textrm{in})$ in Eq.\ (\ref{SCE-gen}) can be naturally split into two --- spectral and kinetic --- contributions:
\be 
\mathcal{F}_\textrm{neq} = \mathcal{F}_\textrm{neq}^\textrm{sp} + \mathcal{F}_\textrm{neq}^\textrm{kin} ,
\ee
where
\begin{subequations}
\label{F&F}
\be
\label{F-neq-sp}
\mathcal{F}_\textrm{neq}^\textrm{sp}
=
- \frac{1}{4\Delta} \int d\epsilon \, F_0(\epsilon)
\Im \left\{ C^R_{\epsilon\epsilon}
\cos\theta^R_\epsilon \sin[\theta^R_\epsilon+\theta^R_{\epsilon-\omega}]
\right\} 
,
\ee
and
\begin{multline}
\label{F-neq-kin}
\mathcal{F}_\textrm{neq}^\textrm{kin}
=
\frac{1}{8\Delta} \int d\epsilon \, D_{\epsilon\epsilon} [F_0(\epsilon)-F_0(\epsilon-\omega)]
\\{}
\times \Im \left\{ \sin\theta^R_{\epsilon-\omega} -
\sin[\theta^R_{\epsilon-\omega}+\theta^R_\epsilon+\theta^A_\epsilon]
\right\} .
\end{multline}
\end{subequations}
Here $C_{\eps\eps}$ and $D_{\eps\eps}$ are the zero-dimensional cooperon and diffuson defined as
\begin{subequations}
\label{CD}
\begin{gather}
C^\alpha_{\epsilon\epsilon}
=
\frac{1}{2\mathcal{E}^{\alpha}_{\epsilon}
+ 2 \Gamma \cos2\theta^\alpha_\epsilon} ,
\\
D_{\epsilon\epsilon}
=
\frac{1}{
\mathcal{E}^{R}_{\epsilon} + \mathcal{E}^{A}_{\epsilon}
- \Gamma [1+\cos(\theta^R_\epsilon-\theta^A_{\epsilon})] \cos(\theta^R_\epsilon+\theta^A_{\epsilon})} ,
\end{gather}
\end{subequations}
where $\alpha=R, A$, and we use the notation $
\mathcal{E}^{R,A}_{\epsilon}
=
\pm (-i\epsilon^{R,A}\cos\theta^{R,A}_\epsilon + \Delta\sin\theta^{R,A}_\epsilon)
$.

The results (\ref{F&F}) allow for a natural interpretation in terms of the microwave-generated correction to the stationary (time-averaged) component of the spectral angle and the stationary (time-averaged) component of the distribution function, correspondingly. Indeed, extracting the linear in $\alpha$ corrections to $\theta^R_\eps$ and $\delta F(\eps)$, we get
\begin{subequations}
\label{first-corr}
\be
\label{dthetaR1}
\delta \theta^R_\eps
= 
- \frac{\alpha}{4}
C^R_{\eps\eps} 
\sin \left( \theta_\eps^R + \theta_{\eps-\omega}^R \right)
+ \{\omega\to-\omega\} 
\ee
and
\begin{multline}
\label{dF1}
\delta F(\eps) 
= 
-
\frac{\alpha D_{\eps\eps}[F(\eps)-F(\eps-\omega)]}{8\cos[(\theta^R_\eps-\theta^A_\eps)/2]} 
\\{}
\times
\left[ 
\cos
\left(\theta_{\eps-\omega}^R + \frac{\theta_\eps^R+\theta_\eps^A}{2} \right)
+ 
\cos
\left(\theta_{\eps-\omega}^A + \frac{\theta_\eps^R+\theta_\eps^A}{2} \right)
\right]
\\{}
+ \{\omega\to-\omega\} .
\end{multline}
\end{subequations}
Substituting now Eqs.\ (\ref{first-corr}) into the \emph{equilibrium}\/ Eq.\ (\ref{F-eq}), we recover the \emph{nonequilibrium}\/ contributions (\ref{F&F}).

\begin{figure}
\centering
\includegraphics[width=0.44\textwidth]{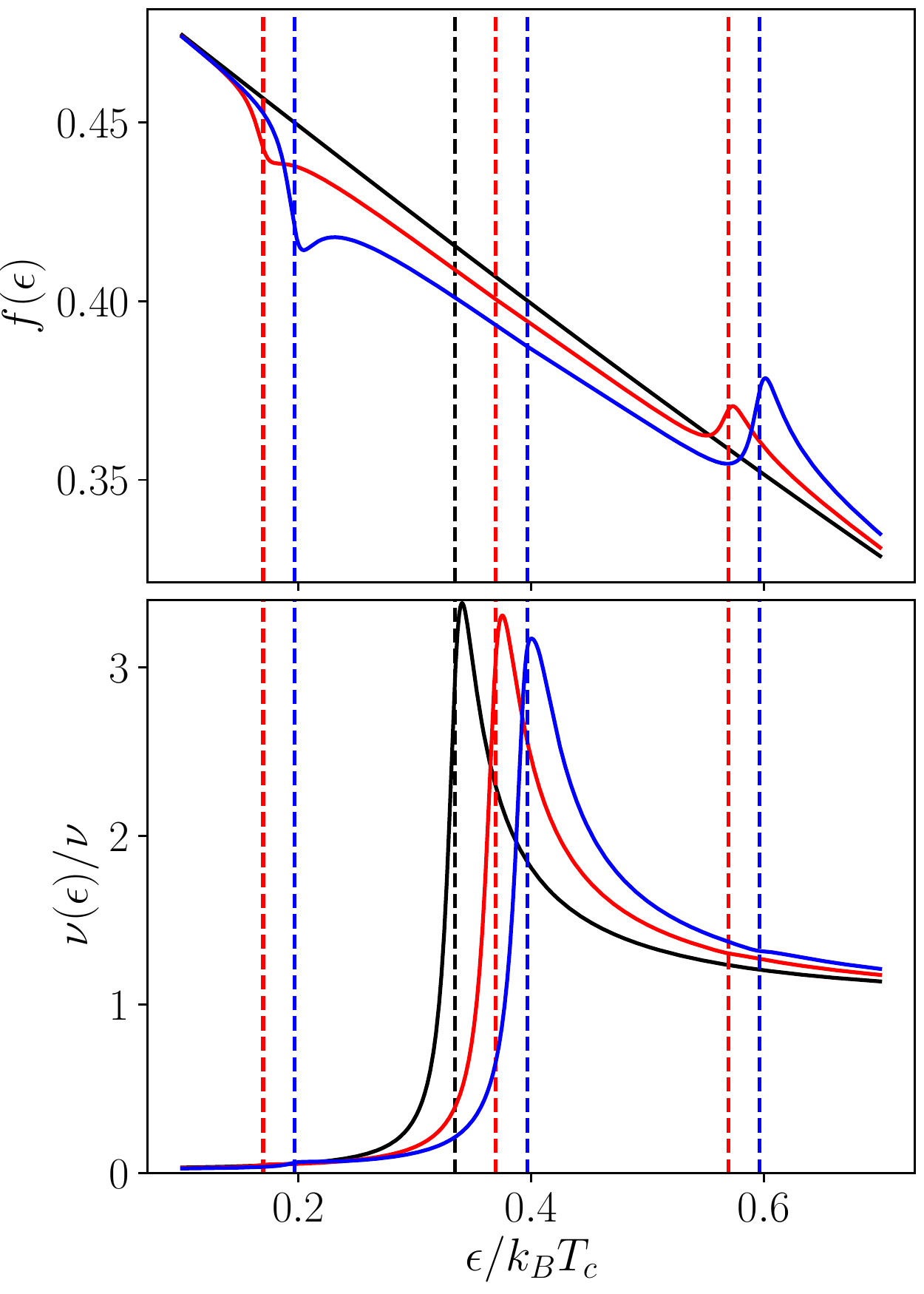}
\caption{Microwave enhancement of superconductivity at $T/T_c=0.98$. Modification of the (time-averaged) (a) quasiparticle distribution function $f(\eps)=[1-F(\eps)]/2$ and (b) density of states.
Black to red: $\alpha/k_BT_c=0$, 0.005, and 0.01. Other parameters: $\hbar\omega=20\gamma_\text{in}$, $\gamma_\text{in}/k_B T_c=0.02$. 
Dashed lines mark the energies $\Delta\pm\hbar\omega$.}
\label{dF-dnu}
\end{figure}

In Fig.\ \ref{dF-dnu}, we illustrate the influence of microwave radiation on the (time-averaged) distribution function $f(E)=[1-F(E)]/2$ and the (time-averaged) DOS $\nu(\eps)/\nu=\Re\cos\theta^R_\eps$ in the GL limit $T\to T_c$. Other parameters are chosen such that microwaves enhance superconductivity (see Fig.\ \ref{wmin-gap}). With increasing the radiation power, the smeared peak in the DOS moves towards larger energies, indicating the growth of $\Delta$.

\begin{figure}
\centering
\includegraphics[width=0.44\textwidth]{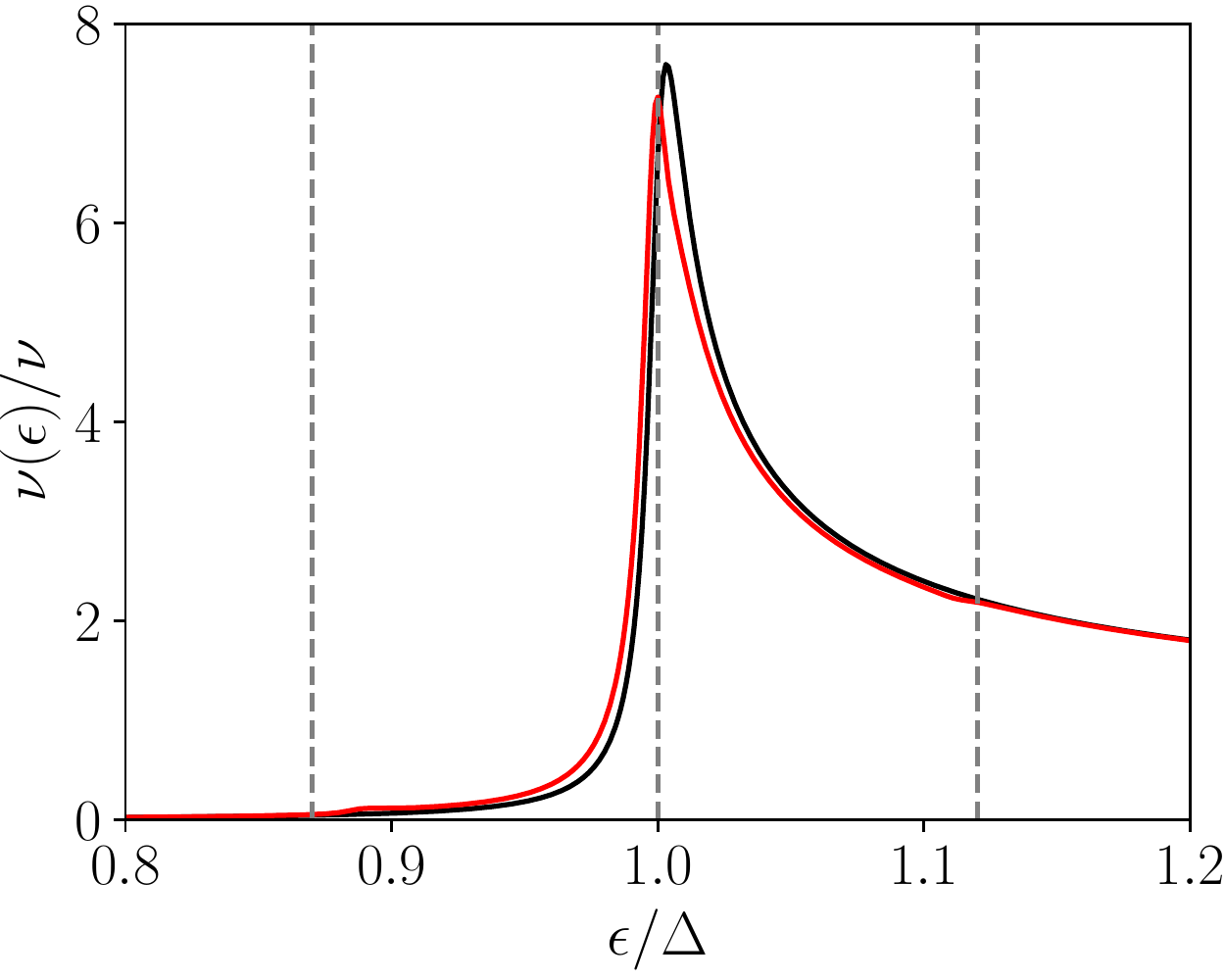}
\caption{Microwave-induced modification of the (time-averaged) density of states at low temperatures ($T\ll T_c$), where the effect of nonequilibrium quasiparticles is negligible and therefore irradiation suppresses superconductivity.
Black to red: $\alpha/k_BT_c=0$ and 0.005.
Other parameters are the same as in Fig.\ \ref{dF-dnu}.}
\label{dnu-lt}
\end{figure}

On the contrary, in Fig.\ \ref{dnu-lt} we plot the (time-averaged) DOS in the low-temperature regime. Here the increase of the microwave power suppresses the spectral gap, in accordance with the phase diagram of Fig.\ \ref{wmin-gap}.

%%%%%%%%%%%%%%%%%%%%%%%%%%

%%%%%%%%%%%%%%%%%%%%%%%%%%

\subsection{Vicinity of $T_c$ and Eliashberg theory}

Another situation, where the effect of microwaves can be treated analytically, is the vicinity of the critical temperature, $T\approx T_c$. In the presence of a \emph{dc} supercurrent, Eq.\ (\ref{SCE-gen}) becomes:
\begin{equation}
\frac{7\zeta(3)}{8\pi^2}\left(\frac{\Delta}{k_{B}T_{c}}\right)^2
-
\frac{T_{c}-T}{T_{c}}
+
\frac{\pi\Gamma}{4k_{B}T_{c}}
=
\alpha \mathcal{F}_\text{neq} ,
\label{GL}
\end{equation}
where the left-hand side is the usual expansion in the absence of radiation (with the last term describing depairing due to a \emph{dc} supercurrent), while the right-hand side perturbatively accounts for the \emph{ac} component of the vector potential. Equation (\ref{GL}) looks very similar to that derived in the Eliashberg theory \cite{eliashberg70} and its earlier generalizations \cite{eckern1979stability,van1984inelastic}. The difference is that our $\mathcal{F}_\text{neq}$ in the right-hand side is a complicated function of $\omega$, $\Delta$, $\Gamma$ and $\gamma_\textrm{in}$, whereas the Eliashberg theory assumed inelastic relaxation to be the slowest process and implied the following set of inequalities:
\be
\gamma_\textrm{in}\ll(\hbar\omega,\Delta)\ll k_B T .
\label{Eliash-cond}
\ee
Under these conditions the function $\mathcal{F}_\text{neq}$ in the right-hand side of Eq.\ (\ref{GL}) acquires the form
\be
\mathcal{F}_\text{neq}
=
- \frac{\pi}{8k_{B}T_{c}}
+
\frac{\hbar\omega}{16\gamma_\textrm{in}k_BT_c} G\left(\frac{\Delta}{\hbar\omega}, \frac{\Gamma}{\Delta}\right) ,
\label{OmegaA}
\ee
where the first term is due to the modification of the static spectral functions (depairing), while the second term has a kinetic origin (quasiparticle redistribution). 

\emph{Gap enhancement.}
In the Eliashberg limit \eqref{Eliash-cond}, the dynamic response of a superconductor in the absence of a \emph{dc} supercurrent ($\Gamma=0$) was calculated in Ref.\ \cite{eckern1979stability}, where the following expression for the function $G_0(\Delta/\hbar\omega)=G(\Delta/\hbar\omega,0)$ was obtained:
\begin{equation}
G_{0}(u)
=
\begin{cases}
2\pi u\left( 1-u^{2}\right)^{-1/2}, & u<1/2 , \\
4 [ K +4u^2 ( \Pi - K ) ]/(2u+1), & u>1/2 ,
\end{cases}
\label{G0}
\end{equation}
where $K=K(k)$ and $\Pi=\Pi(a,k)$ denote complete elliptic integrals of the first and the third kinds \cite{elliptic-comment}, 
and $a=1/(2u+1)^2$, $k=(2u-1)^2/(2u+1)^2$.
Solving Eqs.\ \eqref{GL} and \eqref{OmegaA} with $\Gamma=0$ and $G=G_0(\Delta/\hbar\omega)$ one can find the value of $\Delta(T)$. Comparing it with the equilibrium value of $\Delta_0(T)$ in the absence of the microwave field (in the GL region given by $\Delta_0(T)=\pi k_B[8T_c(T_c-T)/7\zeta(3)]^{1/2}$), one can determine the regions of superconductivity suppression and enhancement. The Eliashberg theory predicts a minimal frequency, $\omega_\text{min}(T)$, for superconductivity enhancement,
which can be obtained from the equation $G_0(\Delta_0(T)/\hbar\omega) = 2\pi\gamma_\text{in}/\hbar\omega$. With $G_0(u)$ given by Eq.\ \eqref{G0}, this equation has the only solution [no upper limit for superconductivity enhancement, $\omega_\text{max}(T)$, see below] given by \cite{eliashberg70,mooij1981enhancement}
\be
\label{wmin-Eliash}
\hbar\omega_\text{min}(T) 
\approx
\sqrt{
\frac{2\pi \gamma_\text{in}\Delta_0(T)}{\ln[\Delta_0(T)/\gamma_\text{in}]} 
}
.
\ee

The minimal frequency is bounded from below by the inelastic relaxation rate:
$\hbar\omega_\textrm{min,min} \sim \gamma_\textrm{in}$. The precise coefficient here cannot be determined within the Eliashberg theory due to the breakdown of the condition \eqref{Eliash-cond}. Nevertheless if we formally apply Eqs.\ \eqref{OmegaA} and \eqref{G0} at the border of their applicability, we obtain
$\hbar\omega_\textrm{min,min} = \sqrt3\gamma_\textrm{in}$, that corresponds to the cusp of $G_0$ at $\hbar\omega_\textrm{min,min}/\Delta=2$ \cite{mooij1981enhancement}.
The exact value of $\omega_\textrm{min,min}$ can be determined with the help of our theory, which does not rely on the smallness of $\gamma_\text{in}$.
In terms of the function $G_0(u)$, a finite value of $\gamma_\textrm{in}$ leads to the rounding of the cusp at $u=1/2$ and the overall suppression of the function. 
As a result, the enhancement effect becomes less pronounced and hence requires a larger frequency to be observable:
\be
\hbar\omega_\textrm{min,min} = 3.23 \, \gamma_\textrm{in},
\label{omega-min}
\ee
corresponding to $\hbar\omega_\textrm{min,min}/\Delta \approx 1.38$. This minimal frequency can be seen in the inset in Fig.~\ref{wmin-gap}. The numerical factor in Eq.\ \eqref{omega-min} is almost 2 times larger than in the above naive estimate from the Eliashberg theory.

In the approximation of Eq.\ (\ref{OmegaA}), there is no upper frequency limit for the superconductivity enhancement. Indeed, the second (kinetic) term in Eq.\ (\ref{OmegaA}) is always positive and according to Eq.\ \eqref{G0} saturates at the level $\mathcal{F}_\text{neq}^\text{kin} = \pi\Delta/8\gamma_\textrm{in}k_BT_c$ at $\hbar\omega\gg\Delta$. Therefore it always wins over the negative $\mathcal{F}_\text{neq}^\text{sp}=-\pi/8k_{B}T_{c}$ in the limit \eqref{Eliash-cond}, indicating the absence of the upper bound $\omega_\text{max}(T)$.
% in the Eliashberg theory. 

In fact, $\omega_\text{max}(T)$ is determined by the heating effect, not accounted for in the approximation of Eq.\ (\ref{OmegaA}), which is written in the lowest order in $\hbar\omega/k_B T$. Including also the quadratic in $\omega/k_B T$ term, we find an additional negative contribution to $\mathcal{F}_\textrm{neq}^\textrm{kin}$ of purely normal origin, such that Eq.\ \eqref{OmegaA} in the limit $\hbar\omega\gg\Delta$ is replaced by 
\be
\mathcal{F}_\text{neq}
=
- \frac{\pi}{8k_{B}T_{c}}
+
  \frac{\pi}{8\gamma_\textrm{in}}
  \left[
  \frac{\Delta}{k_BT_c}
- \frac{7\zeta(3) (\hbar\omega)^2}{\pi^3 (k_B T)^2}
  \right] .
%\label{OmegaA}
\ee
The new contribution establishes an upper bound $\omega_\text{max}$ for the enhancement effect, which remains finite in the limit $\gamma_\text{in}\to0$: %and 
%in the vicinity of $T_c$ 
%is given by 
\be
\label{wmax}
  \hbar \omega_\text{max}(T)
  = 
  1.92\sqrt{\Delta(T) k_BT} 
  \propto
  (1-T/T_c)^{1/4}.
\ee
Note that in the vicinity of $T_c$, $\hbar\omega_\text{max}(T)$ parametrically exceeds the energy scale $2\Delta(T)$, indicating that superconductivity may be enhanced even in the absence   of an obvious gap protection.

\emph{Critical current enhancement.}
In the presence of a supercurrent, the BCS singularity in the DOS gets smeared even in the limit of vanishing $\gamma_\text{in}$. Using the analogy with the Abrikosov-Gor'kov theory of paramagnetic impurities, this smearing can be estimated as $w=(3/2)\Delta\left(\Gamma/\Delta\right)^{2/3}$.
The critical value of the current density in the GL region described by Eq.\ (\ref{GL}) corresponds to $\Gamma_c=4k_B(T_c-T)/{3\pi}$. 
As a result, in the limit $\hbar\omega\ll w\ll \Delta$ the logarithmic integration for $G$ is cut off by $w$ instead of $\hbar\omega$ and the enhancement function $G$ becomes \cite{van1984inelastic} 
\be
\label{G-Son}
G\left(\frac{\Delta}{\hbar\omega}, \frac{\Gamma_c}{\Delta}\right)
=
\frac{2\hbar\omega}{\Delta}\ln\left(9.9\Delta/w\right) .
\ee

The value of the current density in the GL region is determined by 
\begin{equation}
\label{js-1}
j_s/j_0
=
\sqrt{\frac{\Gamma}{2(k_BT_c)^3}}
\int d\epsilon \, W(\epsilon) F(\epsilon) ,
\end{equation}
where
\be
\label{j0}
j_0 = e\nu k_B T_c \sqrt{\frac{D k_B T_c}{\hbar}} ,
\ee
and $\nu$ is the DOS at the Fermi level per one spin projection.
The weight function $W(\epsilon)=\Im \sin^2\theta^R_\epsilon$ in Eq.\ \eqref{js-1} determined by the spectral angle reduces to $W(\epsilon)=\pi\epsilon\delta(|\epsilon|-\Delta)$ for negligible pair breaking and acquires a width $w$ in the presence of a \emph{dc} supercurrent.

%%%%%%%%%%%%%%%%%%%%%
\begin{figure*}[tbp]
\minipage{0.44\textwidth}\includegraphics[width=\textwidth]{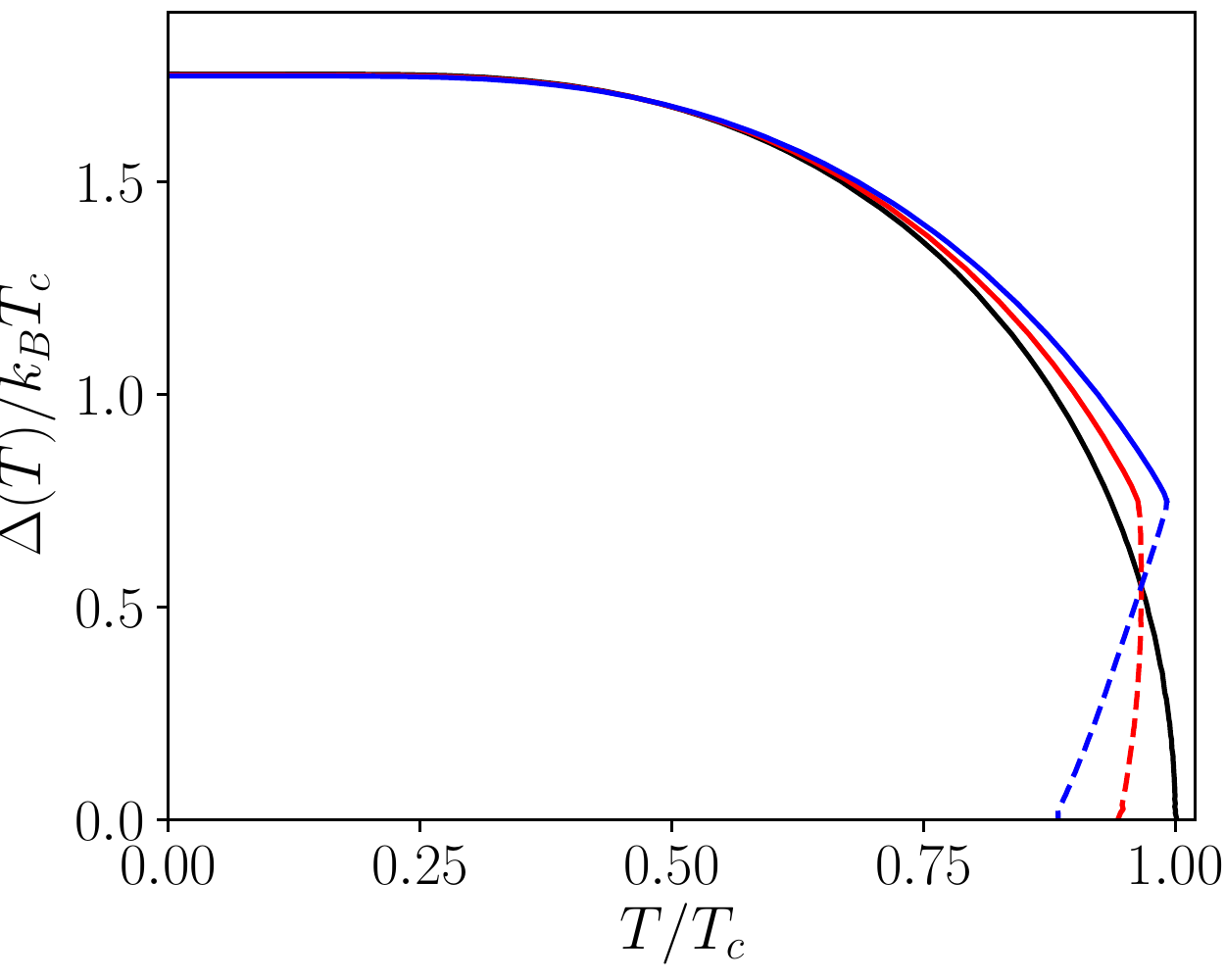}\endminipage
\hspace{0.05\textwidth}
\minipage{0.44\textwidth}\includegraphics[width=\textwidth]{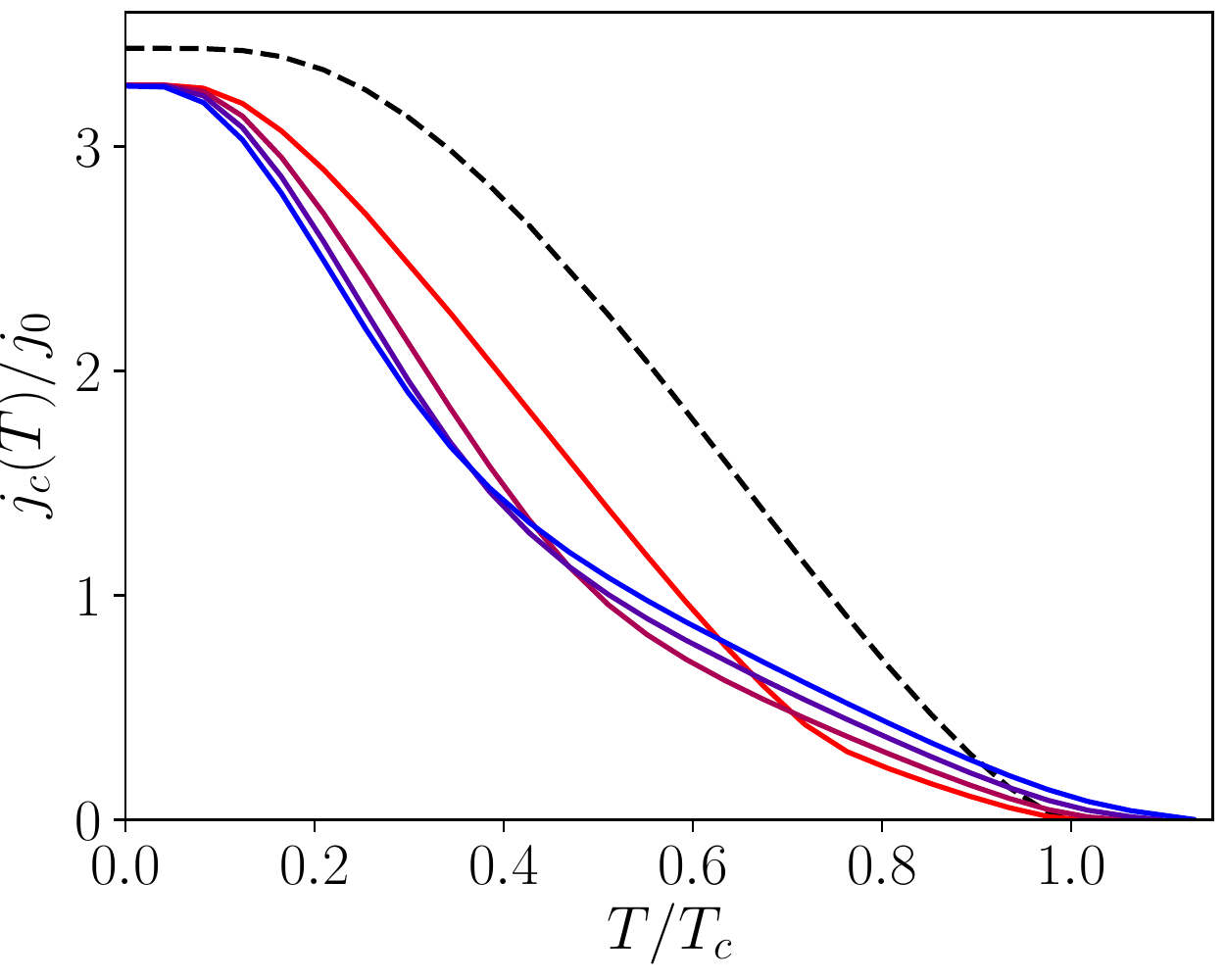}\endminipage
\caption{(a) Temperature dependence of the (time-averaged) order parameter in the absence of a \emph{dc} supercurrent. Microwave power $\alpha/k_BT_c = 0$ (BCS, black), $0.005$ (red) and $0.01$ (blue).
Other parameters: $\hbar\omega/k_BT_c=1.5$ and $\gamma_\text{in}/k_BT_c=0.02$. 
The stable (unstable) branches are shown by solid (dashed) lines.
The sequence of the curves is changed around $0.5\, T_c$, marking a crossover from gap enhancement at high temperatures to gap suppression at low temperatures. 
(b) Temperature dependence of the critical current. Microwave frequency $\omega=0$ (BCS, black dashed), $0.1 \, k_B T_c/\hbar$ to $0.4 \, k_B T_c/\hbar$ (color lines). 
Other parameters: $\alpha=0.05\,k_B T_c$ and $\gamma_\text{in}/k_BT_c=0.02$.}
\label{zero-current:jcall}
\end{figure*}
%%%%%%%%%%%%%%%%%%%%%%%%

\subsection{Full phase diagram}

A typical temperature dependence of the order parameter at zero \emph{dc} supercurrent is shown in Fig.\ \ref{zero-current:jcall}a. At some value of $\alpha>0$, the function $\Delta(T)$ becomes two-valued, with the upper (lower) branch being the stable (unstable) solution
\cite{schmid1977stability,eckern1979stability}. 
Termination of the stable branch (shown by solid lines) marks the actual point of the first-order phase transition in the presence of microwave radiation.
One can see that even if superconductivity is enhanced in the vicinity of $T_c$, this trend turns into superconductivity suppression at lower temperatures. To determine the regions of enhancement/suppression of the order parameter, we consider the limit of weak electromagnetic irradiation ($\alpha\to0$), where the boundary between the these regions is determined from the condition
$\mathcal{F}_\textrm{neq}(\Delta_0(T),0,T,\omega,\gamma_\textrm{in}) = 0 
$
[the order of arguments as in Eq.\ \eqref{SCE-gen}].
For a given value of the inelastic rate $\gamma_\text{in}$, the solution of this equation defines the curve $\mathcal{C}$ in the $(\omega,T)$ plane shown in Fig.~\ref{wmin-gap} for $\gamma_\text{in}/k_BT_c=0.02$. 
In the limit of small $\gamma_\text{in}$, this curve is almost insensitive to $\gamma_\text{in}$, except for the vicinity of the critical temperature, where it marks the lower bound $\omega_\text{min,min}$ for the gap enhancement, as discussed above.
Starting with $\omega_\text{min,min}$ near $T_c$, the lower part of the curve $\mathcal{C}$ describes the evolution of $\omega_\text{min}(T)$ from the GL region, where it is given by Eq.\ \eqref{wmin-Eliash}, to low temperatures. 

The phase diagram in Fig.~\ref{wmin-gap} demonstrates that besides the minimal frequency, $\omega_\text{min}(T)$, there exists a maximal frequency $\omega_\text{max}(T)$ for gap enhancement. Thus the region of stimulated superconductivity encompassed by the curve $\mathcal{C}$ in Fig.~\ref{wmin-gap} is bounded both at low temperatures (no states available) and at high frequencies (heating-dominated regime).
A weak microwave signal always suppresses the superconducting order parameter if the temperature is smaller than $T_\text{min}\approx 0.47 \, T_c$ or the frequency is larger than $\omega_\text{max}\approx 3.3 \, k_BT_c/\hbar$, even though the distribution function continues to have a non-thermal structure. 

In the limit of small temperatures, $T\ll T_c$, the effect of quasiparticle redistribution (kinetic contribution) is negligible because of the gapped DOS, and the main impact of irradiation is modification of the spectral functions \cite{semenov2016coherent}. The spectral contribution to the function $\mathcal{F}_\textrm{neq}^\textrm{sp}$ is given by Eq.\ (\ref{F-neq-sp}). 
In the quasistationary limit, $\omega\ll\Delta$,
one finds $\mathcal{F}_\textrm{neq}^\textrm{sp}=-\pi/8\Delta$, and using $\mathcal{F}_\textrm{eq}=\ln(\Delta/\Delta_0)$ from Eq.\ (\ref{F-eq}), and we obtain for the gap suppression by microwaves: $\Delta=\Delta_0-\pi\alpha/8$. This expression can be readily derived from the Abrikosov-Gor'kov theory \cite{AGmagnetic} with the depairing rate $\alpha/2$ (the factor 1/2 is due to time averaging). In the low-temperature limit it is also possible to calculate the suppression of the superfluid density by radiation using the theory of electromagnetic response of a superconductor with paramagnetic impurities \cite{Skalski}: $\delta n_s/n_s = - (\pi/4+4/3\pi)(\alpha/2\Delta)$, where the first term comes the BCS contribution with the reduced $\Delta$, and the second term is due to modification of the spectral angle. This is equivalent to the modification of the kinetic inductance: $\delta L_K/L_K = -\delta n_s/n_s$, as discussed in Refs.\ \cite{semenov2016coherent,Semenov-2B}.

The theory developed in Ref.\ \cite{Tikhonov2018} also allows for determination of the microwave effect on the critical current $j_c(T)$ in a superconductor. The latter should be obtained by maximization of the current density $j_s$ with respect to the order parameter $\Delta$. The results for $j_c(T)$ are shown in Fig.\ \ref{zero-current:jcall}b for several frequencies at a fixed microwave power. The critical current at equilibrium \cite{kupryanov1980temperature,romijn1982critical} is shown by the dashed line. One can clearly see the difference between the high- and low-temperature regions. At high temperatures, $T\sim T_c$, the critical current is enhanced by microwaves, but the frequency required to its enhancement grows with the temperature decrease, consistent with previous studies. However at low temperatures the trend reverses to the opposite: the larger is the frequency, the stronger is the critical current suppression. Physically this behavior originates from freezing out of the kinetic contribution, while the effect of irradiation on the spectral properties always leads to superconductivity suppression via the pair-breaking mechanism.

The region on the phase diagram in Fig.~\ref{wmin-gap} where the critical current is enhanced by a weak microwave field is encompassed by the curve $\mathcal{C}'$. This region is a subset of the gap enhancement region shown by the curve $\mathcal{C}'$.

\section{Conclusion}

We have studied behavior of a dirty superconducting wire, which may carry a \emph{dc} supercurrent, under weak \emph{ac} electromagnetic driving, generalizing the Eliashberg theory \cite{eliashberg70,ivlev1971influence} to higher driving frequencies, lower temperatures and finite supercurrent density. The most important feature of our theory is that the effect of quasiparticle redistribution is treated on equal footing with the modification of the spectral properties. Physically, our results are determined by the interplay between several competing effects of the microwaves: 
(i)~non-equilibrium redistribution of quasiparticles with sub-thermal features responsible for stimulation of superconductivity, 
(ii)~Joule heating, and 
(iii) modification of the spectral functions due to depairing. The resulting phase diagram is shown in Fig.\ \ref{wmin-gap}, where the criteria for the microwave-stimulated enhancement (a) of the gap and (b) of the critical current are presented. We reveal that the gap enhancement is observed in a finite region of the $(\omega,T)$ plane, roughly limited by the conditions $T>0.5 \, T_c$ and $\hbar\omega<3 \, k_B T_c$. 
The absence of the gap enhancement at low $T$ is due to the suppression of available quasiparticle DOS switching off the mechanism (i), whereas at large frequencies, the dominant effect is the Joule heating (ii).
In the presence of a \emph{dc} supercurrent, the role of the mechanism (iii) is increased that makes the region of the critical current enhancement narrower than the region of the gap enhancement.

Following the Eliashberg theory, our approach relies on the assumption of spatial homogeneity, when both the absolute value and the phase gradient of the order parameter are the same at every point in the wire. Then gauging out the phase one arrives at a zero-dimensional problem to be solved. Spontaneous breakdown of the translational symmetry leading to inhomogeneous non-equilibrium states was investigated in the framework of the Eliashberg theory in Ref.\ \cite{eckern1979stability}. It remains an open problem to study this effect for lower temperatures.

One of the most straightforward applications of the developed theory is the devices based on superconducting microresonators, for instance, Microwave Kinetic Inductance Detectors (MKID) which have been shown to be promising for astronomical studies \cite{Day2003,zmuidzinas2012superconducting,baselmans2017kilo}. In order to achieve a sufficiently high signal-to-noise ratio, given the existing low noise amplifiers, the microwave read-out signal is increased to a regime where a significant effect on the superconducting properties is observed. This has recently driven the study of the microwave response of superconductors at low temperatures \cite{de2014fluctuations,de2014evidence,sherman2015higgs,moor2017amplitude}. 
Our theoretical predictions can be used to analyze measurements on MKID \cite{de2014fluctuations,de2014evidence}, as well as in the experiment proposed in Ref. \cite{Semenov-2B}. Apart from that, there are many controllable ways to drive superconducting systems out-of-equilibrium: disturbing them by a supercritical current pulse \cite{geier1982response,frank1983transient}, imposing to pulsed microwave phonons \cite{tredwell1975phonon},
or directly injecting non-equilibrium quasiparticles \cite{van1987enhancement1,van1987enhancement2}. It would be interesting to study these problems microscopically in the similar framework.

\acknowledgments

The authors are grateful to T. M. Klapwijk for stimulating our interest in low-temperature generalization of the Eliashberg theory.
The research of AS and KT was supported by the Russian Science Foundation, Grant No.\ 17-72-30036.
The research of MS was partially supported by Skoltech NGP Program (Skoltech-MIT joint project).

\bibliographystyle{apsrev}
\bibliography{scenh}

\begin{thebibliography}{58}
\bibitem[{gam()}]{gamma-com}
\bibinfo{note}{Our $\gamma_\text{in}$ is two times larger than $\gamma$ defined
  in the original Eliashberg's paper [10].}
\bibitem[{ell()}]{elliptic-comment}
\bibinfo{note}{We follow Wolfram Mathematica notations for elliptic integrals.}
\end{thebibliography}

\end{document}